\documentclass[twocolumn,twocolappendix]{aastex701}

\usepackage{amsmath}

\newcommand{\F}{\mathcal{F}}
\newcommand{\asini}{a_{\mathrm{p}}}

\def\tinytt#1{\tiny{\textrm{#1}}}

\newcommand\Tstrut{\rule{0pt}{2.9ex}}       
\newcommand\Bstrut{\rule[-1.3ex]{0pt}{0pt}} 
\newcommand\TBstrut{\Tstrut\Bstrut}

\usepackage{soul}

\newcommand{\hvalue}{2.63 \times 10^{-25}}
\newcommand{\hfreq}{211.1}
\newcommand{\evalue}{5.2 \times 10^{-8}}

\newcommand{\avalue}{1.5 \times 10^{-6}}

\graphicspath{ {./Plots/} {./Figures/} {./figs/} {./} }
\DeclareGraphicsExtensions{.png}

\begin{document}

\title{Wide parameter-space O3 search for continuous gravitational waves from unknown neutron stars in binary systems}

\author[0000-0002-1845-9309]{P.~B.~Covas}
\affiliation{Max Planck Institute for Gravitational Physics (Albert Einstein Institute) and Leibniz Universit\"at Hannover,\\
D-30167 Hannover, Germany}
\affiliation{Departament de Física, Universitat de les Illes Balears, IAC3, Carretera Valldemossa km 7.5, E-07122 Palma, Spain}
\email{jb.covas@uib.es}
\author[0000-0002-1007-5298]{M.~A.~Papa}
\affiliation{Max Planck Institute for Gravitational Physics (Albert Einstein Institute) and Leibniz Universit\"at Hannover,\\
D-30167 Hannover, Germany}
\email{maria.alessandra.papa@aei.mpg.de}
\author[0000-0002-3789-6424]{R.~Prix}
\affiliation{Max Planck Institute for Gravitational Physics (Albert Einstein Institute) and Leibniz Universit\"at Hannover,\\
D-30167 Hannover, Germany}
\email{reinhard.prix@aei.mpg.de}


\begin{abstract}
Continuous gravitational waves, i.e., persistent and nearly-monochromatic signals emitted by
asymmetric spinning neutron stars, remain elusive.
Searches for these signals from unknown {\it{binary}} systems are the most computationally challenging, but
they are essential, given that binary accretion provides a natural mechanism for creating the
required asymmetry, and around half of the known pulsars rotating above 25\,Hz are part of a binary
system.
Here we report on a search of a large uncharted parameter-space region: for the
first time we cover gravitational-wave frequencies above 520\,Hz (from 50 to 1\,000\,Hz), and, for the
first time with advanced detectors, orbital periods lower than 3\,days are explored.
No signal is detected, and we set the most stringent constraints to date on the amplitude of signals of this kind. 
Our results exclude with $95\%$ confidence neutron stars within 100\,pc and rotating faster than 
$\sim$ 495 Hz from having ellipticities above $\evalue$. 
Within the same distance our results also exclude r-mode amplitudes above 
$\avalue$ for stars rotating faster than $\sim$ 740 Hz.
\end{abstract}


\section{Introduction}
\label{sec:intro}

Continuous gravitational waves (CWs), which are yet to be detected, may be sourced by rotating neutron stars with a time-varying quadrupole \citep[][]{KeithReview}.
This quadrupole can be generated by a mass asymmetry or an unstable oscillation mode like the r-mode \citep[][]{jones2024multimessengerobservationsscienceenabled}.

Searches for these signals can be categorized based on the type of prior knowledge of the source.
Targeted searches focus on known pulsars, while all-sky searches scan for entirely unknown objects, 
which is crucial since most galactic neutron stars remain undiscovered.

Due to their large parameter spaces, fully-coherent all-sky searches are computationally 
prohibitive \citep{WETTE2023102880} and hence semi-coherent methods \citep[][]{Tenorio_2021} 
are usually employed. Semi-coherent methods split the data into shorter segments that are 
searched coherently and then incoherently combine together the results. Semi-coherent searches 
are less sensitive than coherent searches on the same data, but they can be tuned to be computationally feasible.
Promising candidates from an initial semi-coherent search can then be further assessed 
with more sensitive semi-coherent searches having increasingly longer baseline segments.
 
The computational challenge increases when searching for unknown neutron 
stars in binary systems, as their orbital motion introduces additional parameters \citep{Leaci:2015bka}.
For this reason only a relatively small subset of the possible signals from neutron 
stars in binaries have been explored. It is important to expand the searches for continuous 
wave emission from neutron stars in binaries because a large fraction of known millisecond 
pulsars are in binaries \citep{Manchester_2005}, and because accretion may have generated the quadrupole moment \citep{mountainacc,10.1093/mnras/stad967,Singh:2019dgy} required to make them detectable through their continuous gravitational-wave emission.

\begin{figure*}
  \begin{center}
    \includegraphics[width=0.49\textwidth]{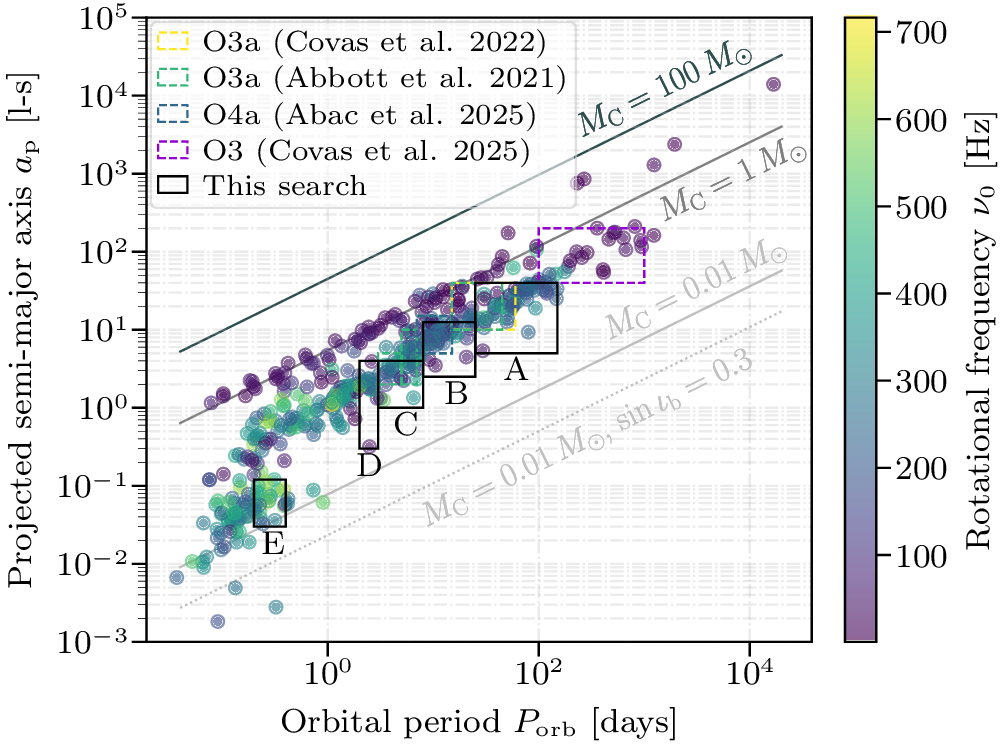}
    \includegraphics[width=0.49\textwidth]{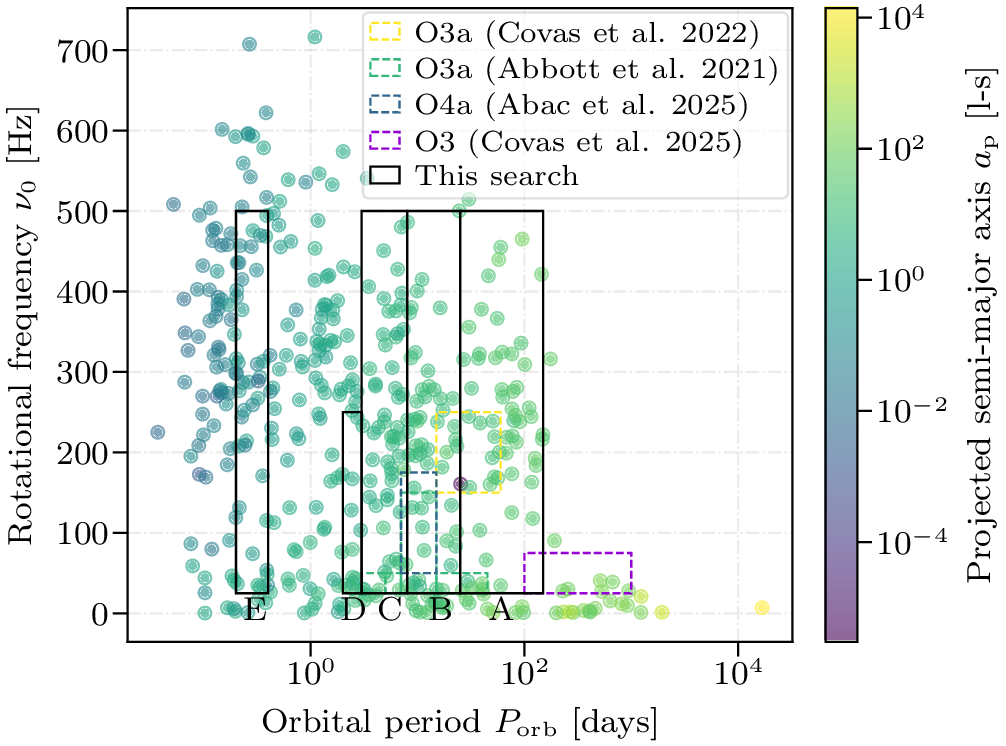}
    \caption{The black boxes show the signal-parameter regions covered by this search (left plot shows $\{ P_\mathrm{orb}, a_{\mathrm{p}} \}$, right plot shows $\{ P_\mathrm{orb}, \nu_0 \}$).
    The circles show the known pulsars in binary systems from the ATNF catalog version 2.6.3 \citep{Manchester_2005} and the color encodes $\nu_0$ (left plot) and $a_{\mathrm{p}}$ (right plot).
    The dashed boxes show the ranges covered by previous searches using O3 or O4 Advanced LIGO data.
    The different gray lines on the left plot show Kepler's third law for different values of the companion mass $M_\mathrm{C}$ and inclination of the binary orbit $\iota_b$ (continuous lines assume $\iota_b = 90\deg$), assuming a neutron star mass of 1.4 M$_{\odot}$.}
    \label{fig:ParameterSpace}
  \end{center}
\end{figure*}

In this paper we present a search using the public O3 data from the Advanced LIGO detectors \citep{GWOSC}.
Our search covers signal 
frequencies up to $1\,000$ Hz (see figure~\ref{fig:ParameterSpace}), almost doubling the maximum frequency previously investigated by a search of this kind.
Allowing for the spin frequency of the objects to be high is important due to the recycling scenario, where neutron stars gain angular momentum due to the accretion of matter from their companion \citep{Patruno_2020}. Furthermore, some studies suggest that gravitational-wave emission could help to explain the observed spins of the galactic pulsar population \citep{Gittins_2019}, which is particularly important for neutron stars with high rotational frequencies like the ones targeted by this search.

\section{The Search}
\label{sec:method}

\subsection{Data}
\label{sec:data}

This search employs publicly available data from the third observing run (O3) \citep{GWOSC} of the Advanced LIGO Hanford (H1) and Advanced LIGO Livingston (L1) gravitational-wave detectors \citep{PhysRevD.102.062003}. 
We use the \texttt{GWOSC-16KHZ\_R1\_STRAIN} channel and the \texttt{DCS-CALIB\_STRAIN\_CLEAN\_SUB60HZ\_C01\_AR} frame type, which is cleaned of non-Gaussian noise at the 60\,Hz harmonics and calibration lines \citep{PhysRevD.97.082002} prior to being publicly released. We further apply the gating procedure of \cite{gating} to remove short-duration glitches that increase the noise floor and reduce search sensitivity \citep[e.g.,][]{O3allskyLIGO}.

\begin{figure}
  \begin{center}
    \includegraphics[width=1.0\columnwidth]{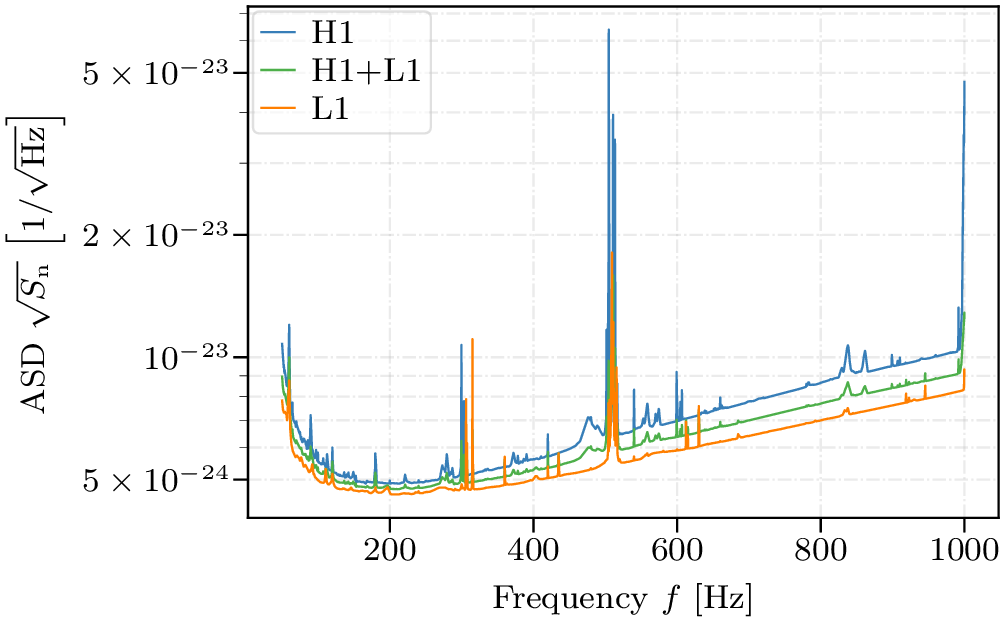}
    \caption{Square root of the harmonic mean power spectral density $\sqrt{S_\mathrm{n}}$ of the data used in this search as a function of frequency. The upper blue curve shows H1, the middle green curve shows H1+L1, and the lower orange curve shows L1.}
    \label{fig:PSD}
  \end{center}
\end{figure}

From this data we produce short-baseline Fourier transforms \citep[SFTs, see][]{Allen_Mendell} of 200\,s ($105\,111$ SFTs for H1 and $105\,269$ SFTs for L1), short enough to contain the signal power in a single frequency bin during this time.
The amplitude spectral density (ASD) $\sqrt{S_\mathrm{n}}$ in the frequency range covered by this search is shown in figure~\ref{fig:PSD}.

\subsection{Signal model and parameter space}
\label{sec:signal}

We assume the CW signal model of \citet{jaranowski_data_1998}, limiting it to emission at a single frequency.

This model depends on four amplitude parameters: the intrinsic gravitational-wave amplitude $h_0$, the angle between the angular momentum of the neutron star and the line of sight $\cos\iota$, the polarization angle $\psi$, and the initial phase $\phi_0$.
The model also depends on several phase-evolution parameters: the intrinsic gravitational-wave frequency $f_0$ and its derivatives $f_1,...,f_k$, the sky position of the source $\alpha, \delta$, and the binary orbital parameters. The latter are the orbital period $P_\mathrm{orb}$, the projected semi-major axis $a_{\mathrm{p}}$, the time of ascension $t_{\mathrm{asc}}$, the argument of periapsis $\omega$, and the eccentricity $e$, following the small-eccentricity model of \citet[][]{Leaci:2015bka} that defines the orbit of the neutron star around the binary barycenter.

We search for signals across the entire sky, with gravitational-wave frequencies $f_0$ indicated in table~\ref{tab:region}, 
covering over 86\% of the known pulsars with orbital parameters in the examined ranges \citep{Manchester_2005}.

We assume 
\begin{equation}
|f_{1}| \le
\begin{cases}
4.67 \times 10^{-11}\,\mathrm{Hz/s} & \text{if } f_0 < 500 \,\mathrm{Hz},\\
6.61 \times 10^{-11}\,\mathrm{Hz/s} & \text{otherwise},
\end{cases}
\label{eq:f1max}
\end{equation}
which is consistent with the observed population of neutron stars in binaries \citep{Manchester_2005} and allows us to set $f_1 = 0\,\mathrm{Hz/s}$ for all templates in the initial stage of the search without increasing the mean mismatch by more than $2\%$ with respect to templates with perfect match in signal spin-down. This choice limits the detectable amplitude to lie below the grey shaded regions of figure~\ref{fig:UL}, which we consider an acceptable trade-off based on the null results of all previous all-sky searches.

The projected semi-major axis $\asini$ of neutron stars in binary systems spans a broad range, determined by the companion mass $M_C$, orbital inclination $\iota_b$, and orbital period $P_\mathrm{orb}$, as illustrated in figure~\ref{fig:ParameterSpace}. Due to the prohibitive computational cost of these searches, we select five different regions of the orbital parameter-space 
$\{ P_\mathrm{orb}, a_{\mathrm{p}} \}$ where neutron stars in binary systems lie (indicated by the black boxes in that figure and given in table~\ref{tab:region}). We also show the range of signal frequencies that we investigate, expressed in terms of the corresponding spin frequency $\nu_0=f_0/2$.  All our search regions contain unexplored areas, and region E was completely unexplored prior to this search \citep[][]{TwoSpectS6,O2aallskybinary,O3aallskybinaryLIGO,Covas_2022,Covas_2025,theligoscientificcollaboration2025allskysearchcontinuousgravitationalwave}.

Neutron stars in region E might be part of low mass X-ray binaries (LMXBs) with a less compact donor star. As shown in the right plot of figure \ref{fig:ParameterSpace}, these neutron stars possess the highest rotational frequencies, which can be explained by the recycling scenario, where the neutron star accretes matter from its companion and gains angular momentum, a procedure which might provide the required asymmetry \citep{mountainacc,10.1093/mnras/stad967,Singh:2019dgy}.
Furthermore, some studies suggest that long-lasting r-modes might be present in quiescent neutron stars in LMXBs, belonging to the so-called HOFNAR population \citep{Chugunov:2014cwu}.
On the other hand, binary systems with a small orbital period are particularly difficult to search due to the increased template density \citep[proportional to $a_{\mathrm{p}}^3 / P_\mathrm{orb}^5$, see equation 77 of][]{Leaci:2015bka} in that region.

We assume that the orbital eccentricity $e$ is small:
\begin{equation}
\label{eq:eccmax}
e \le \min\left(0.1, \hat{e}\right),
\end{equation}
with
\begin{linenomath*}
\begin{equation}
\begin{aligned}
  \hat{e} &\equiv \begin{cases}
    0.073\,\left({200\,\textrm{Hz}\over{f_0}}\right) \left({{P_\mathrm{orb}} \over 10\,\textrm{days}}\right) \left({10\,\textrm{l-s}\over{a_{\mathrm{p}}}}\right) & \text{if } f_0 < 500 \,\mathrm{Hz},\\
    0.026\,\left({800\,\textrm{Hz}\over{f_0}}\right) \left({{P_\mathrm{orb}} \over 10\,\textrm{days}}\right) \left({10\,\textrm{l-s}\over{a_{\mathrm{p}}}}\right) & \text{otherwise},
  \end{cases} \\
\end{aligned}
\label{eq:defeccmax}
\end{equation}
\end{linenomath*}
obtained from equation 36 of \citet{BSHFstat} with maximum mismatch $\mu_M=1.5$ below 500 Hz and $\mu_M=3$ elsewhere.
This allows us to set $e = 0$ (and $\omega = 0$) for all templates in the initial stage of the search without increasing the mean mismatch by more than $2\%$ with respect to templates with perfect match in signal eccentricity (see appendix \ref{sec:robustness} for the Monte-Carlo set-up used to test this). Only 9 out of 152 pulsars in binary systems located in these regions \citep{Manchester_2005} have eccentricities outside of the range defined by equations~\ref{eq:eccmax} and \ref{eq:defeccmax}.

\begin{deluxetable*}{lccccc}
\label{tab:region}
\tablecaption{Range of values for the different parameters covered by the search. $t_m$ is the mid-time of the search, equal to $1\,253\,764\,659.5$ in GPS.
A single value indicates an equal range for all orbital parameter-space regions.}
\tablehead{\colhead{Parameter} & \colhead{Region A} & \colhead{Region B} & \colhead{Region C} & \colhead{Region D} & \colhead{Region E}}
\startdata
\TBstrut $f_0$ [Hz] & [50, 1000] & [50, 1000] & [50, 1000] & [50, 500] & [50, 1000] \\
\TBstrut $|{f_1}|$ [Hz/s] & \multicolumn{5}{c}{see equation \ref{eq:f1max} and its discussion} \\
\TBstrut $\alpha$ [rad] & \multicolumn{5}{c}{[0, $2\pi$)} \\
\TBstrut $\delta$ [rad] & \multicolumn{5}{c}{[$-\pi/2$, $\pi/2$]} \\
\TBstrut $a_{\mathrm{p}}$ [l-s] & [5, 40] & [2.5, 12.5] & [1, 4] & [0.3, 4] & [0.03, 0.12] \\
\TBstrut $P_\mathrm{orb}$ [days] & [25, 150] & [8, 25] & [3, 8] & [2, 3] & [0.2, 0.4] \\
\TBstrut $t_{\mathrm{asc}}$ [GPS] & \multicolumn{5}{c}{[$t_m - P_\mathrm{orb}/2$, $t_m + P_\mathrm{orb}/2$)} \\
\TBstrut $e$ & \multicolumn{5}{c}{see equation \ref{eq:eccmax} and its discussion} \\
\TBstrut $\omega$ [rad] & \multicolumn{5}{c}{[0, $2\pi$)} \\
\enddata
\end{deluxetable*}

\subsection{Initial search}
\label{sec:results}

We use the \textsc{BinarySkyHou$\mathcal{F}$} \citep{BSHFstat} semi-coherent search method with $N_{\mathrm{seg}}=31\,675$ segments of length $T_{\mathrm{seg}} = 900$\,s. 

This search method uses templates with $P_\mathrm{orb}=0$ at the coherent stage, and calculates the $\mathcal{F}_{\mathrm{AB};\ell}$ detection statistic of \cite{2DOF} with $\ell=1, \dots, N_{\mathrm{seg}}$ over a grid in $f_0$ and sky.

We take a linear combination of the $\mathcal{F}_{\mathrm{AB};\ell}$, each tracking the time-frequency pattern produced by a possible signal, now including templates with non-zero values of $\{ P_\mathrm{orb}, a_{\mathrm{p}}, t_{\mathrm{asc}} \}$:
\begin{equation}
\label{eq:twoFsum}
  2 \hat{\mathcal{F}}_{\mathrm{ABw}} \equiv \sum_{\ell=1}^{M} w_\ell \,2\mathcal{F}_{\mathrm{AB};\ell}, 
\end{equation}
with the value of $M$
explained below. The sky-position-dependent weights $w_\ell$ are given by equation 54 of \citet{Prix_2025}. The template grid spacings for the initial stage search are shown in table~\ref{tab:setup}.

\begin{deluxetable*}{lcc}
\label{tab:setup}
\tablecaption{Grid spacings of the initial stage of the search. $\Omega=2\pi/P_\mathrm{orb}$ is the average angular orbital velocity.}
\tablehead{Parameter & $f_0 < 500$ Hz & $ f_0 \ge 500$ Hz }
\startdata
\TBstrut $\delta f_0$ [mHz] & $1.1$ & $1.1$ \\
\TBstrut $\delta \alpha$ [mrad] & $79.4 \, \left({{200 \, \tinytt{Hz}}\over{f_0}}\right)$ & $27.8 \, \left({{800 \, \tinytt{Hz}}\over{f_0}}\right)$ \\
\TBstrut $\delta \delta$ [mrad] & $79.4 \, \left({{200 \, \tinytt{Hz}}\over{f_0}}\right)$ & $27.8 \, \left({{800 \, \tinytt{Hz}}\over{f_0}}\right)$ \\
\TBstrut $\delta \asini$ [l-s] & $ 1.5 \, \left({{200 \, \tinytt{Hz}}\over{f_0}}\right) \left({{P_\mathrm{orb} \over {10 \, \tinytt{days}}}}\right)$ & $ 0.5 \, \left({{800 \, \tinytt{Hz}}\over{f_0}}\right) \left({{P_\mathrm{orb} \over {10 \, \tinytt{days}}}}\right)$ \\
\TBstrut $\delta \Omega$ [nHz] & $ 16.0 \, \left({{200 \, \tinytt{Hz}}\over{f_0}}\right) \left({{P_\mathrm{orb} \over {10 \, \tinytt{days}}}}\right) \left({{10 \, \tinytt{l-s}}\over{\asini}}\right)$ & $ 5.7 \, \left({{800 \, \tinytt{Hz}}\over{f_0}}\right) \left({{P_\mathrm{orb} \over {10 \, \tinytt{days}}}}\right) \left({{10 \, \tinytt{l-s}}\over{\asini}}\right)$ \\
\TBstrut $\delta t_{\mathrm{asc}}$ [kGPS] & $ 20.1 \, \left({{200 \, \tinytt{Hz}}\over{f_0}}\right) \left({{P_\mathrm{orb} \over {10 \, \tinytt{days}}}}\right)^2 \left({{10 \, \tinytt{l-s}}\over{\asini}}\right)$ & $ 7.1 \, \left({{800 \, \tinytt{Hz}}\over{f_0}}\right) \left({{P_\mathrm{orb} \over {10 \, \tinytt{days}}}}\right)^2 \left({{10 \, \tinytt{l-s}}\over{\asini}}\right)$ \\
\enddata
\end{deluxetable*}

In the first stage of the analysis $M=
{\textrm{round}}[{0.75 \times N_{\textrm{seg}}}]$, because we exclude from the sum the segments within the lowest 25th percentile of the $w_\ell$ (what segments are excluded varies with sky position). This reduces the computational cost of the search by $\sim 25\%$ while incurring a loss in sensitivity $\leq$ 5\%, as verified through direct measurements on fake signals added to the data shown in figure \ref{fig:sensdepth}.

The top $1\%$ templates of each 0.1 Hz band are then re-analyzed \citep[as explained in section 4B of][]{BSHFstat} using all segments: $M=N_{\textrm{seg}}$.

We define a significance $s$ for every template as
\begin{equation}
  s \equiv \frac{ 2 \hat{\mathcal{F}}_{\mathrm{ABw}} - \mu}{\sigma},
\label{eq:sig}
\end{equation}
where $\mu \equiv 2\sum_\ell w_\ell = 2 M$ and
$\sigma^2 \equiv 4 \sum_\ell w^2_\ell$ are the expected mean and variance
of $2 \hat{\mathcal{F}}_{\mathrm{ABw}}$ in Gaussian noise.
The significance $s_i$ is our ranking statistic, where the index $i$ indicates the stage of the search, with 0 being the initial stage. 

\begin{figure}
  \begin{center}
    \includegraphics[width=1.0\columnwidth]{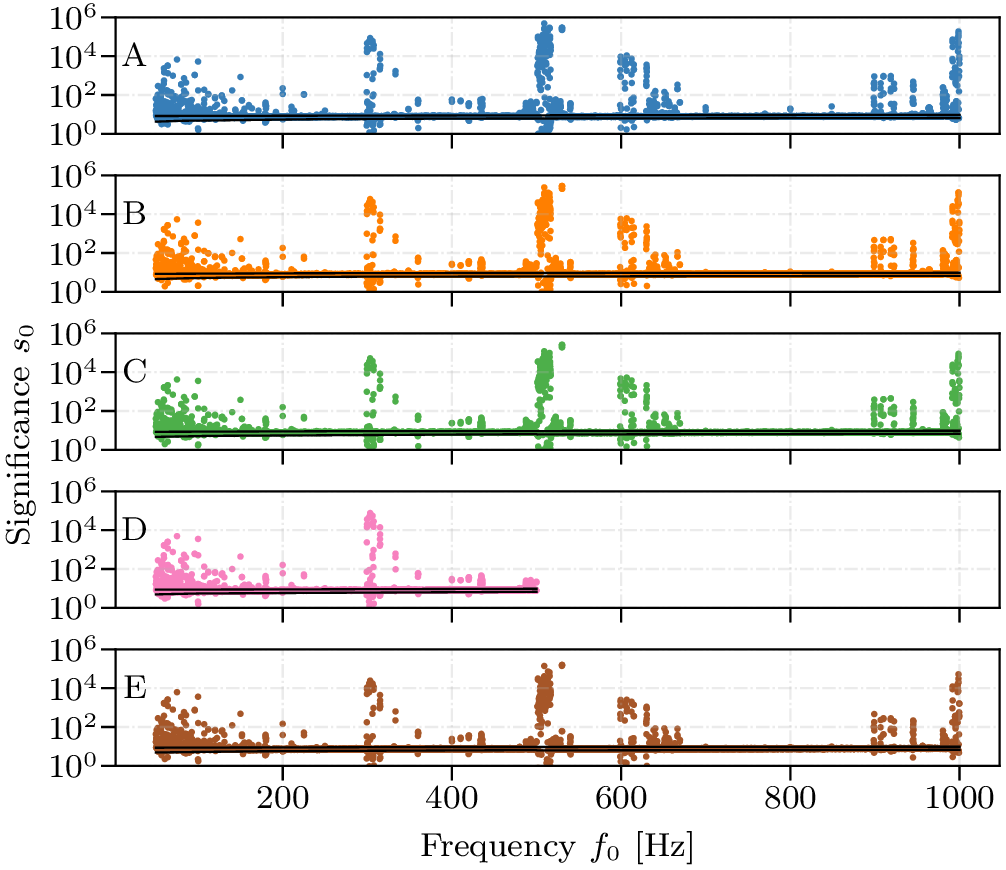}
    \caption{Maximum significance $s_0$ of the initial stage in each 0.1\,Hz frequency band as a function of the gravitational-wave frequency.
    The black lines show the threshold to consider a band non-Gaussian.
    A few bands with negative maximum $s_0$ values are not shown on the plot.}
    \label{fig:PSD2}
  \end{center}
\end{figure}

Our search has an average mismatch of $\sim 70\%$. More details about the set-up of the search are given in appendix~\ref{sec:details}.

Figure~\ref{fig:PSD2} shows the most significant raw results of the search: in each of the five orbital parameter search regions we show the most significant (highest $s_0$) result from every 0.1 Hz band. Prominent features below and above the average level can be attributed to different noise sources such as the 60\,Hz power mains. 

We use the clustering procedure of \citet{2019PhRvD..99l4019C,O2aallskybinary} on initial stage results and group together those likely caused by the same underlying source.
The 3 clusters with the highest significance per 0.1\,Hz band are saved for further analysis as candidates, yielding a total of 127\,500 candidates: 28\,500 candidates from each of the A, B, C, and E regions and 13\,500 candidates from the D region because of its smaller frequency range.

\subsection{Follow-up}
\label{sec:fu}

We follow-up these selected candidates by searching a volume of parameter space around each of them with a coherent time $T_{\textrm{seg}}=2\,700$ s, three times longer than that of the original search. On these reduced volumes we can leverage a nested sampling algorithm \citep{bilby} to calculate the detection statistic \citep{covas2024new}.
We now explicitly search the $f_1$, $e$, and $\omega$ ranges given in table~\ref{tab:region} due to the increased resolution.

To distinguish potential gravitational-wave signals from noise, we require the evolution of a candidate's detection statistic to be consistent with the behavior of an astrophysical signal.
This expected behavior is measured on simulated test-signals added to the O3 data (see appendix~\ref{sec:robustness} for their distribution). In particular we consider two aspects: the consistent increase in significance $s_1(s_0)$ and in the robust detection statistic $\log_{10} \hat{B}_\mathrm{1;S/GLtL}(s_0)$ \citep{PhysRevD.93.084024} across stages 0 and 1. To calculate the $\log_{10} \hat{B}_\mathrm{1;S/GLtL}$ statistic we set the free parameter $\hat{\mathcal{F}}^* = 0$, which is equivalent to comparing a signal hypothesis with a single-detector line hypothesis, thus complementing the standard $\mathcal{F}$-statistic that compares signal and Gaussian noise hypotheses. 

\begin{figure*}
  \begin{center}
    \includegraphics[width=1\columnwidth]{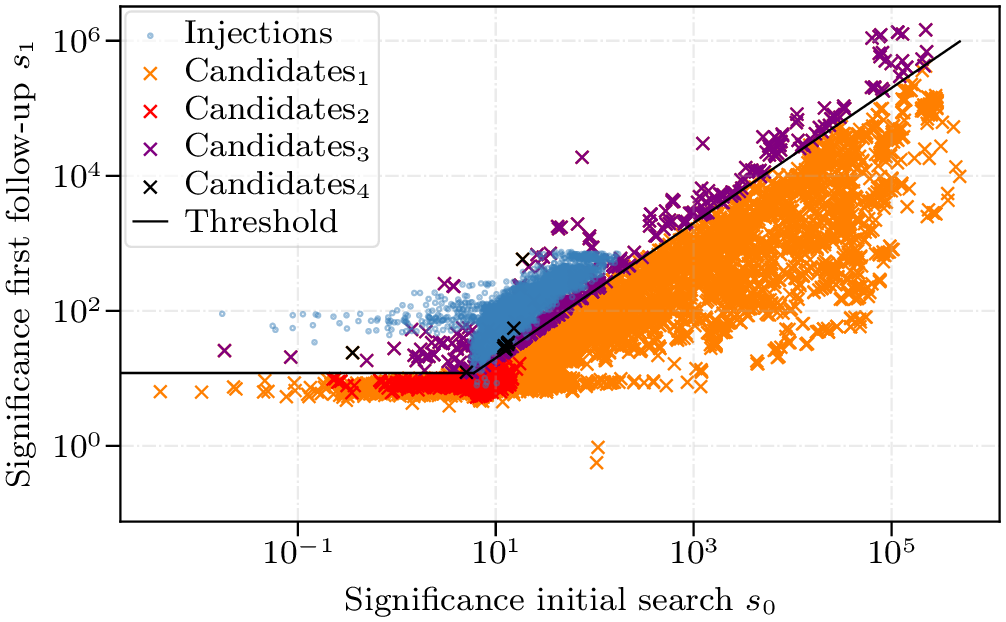}
    \includegraphics[width=1\columnwidth]{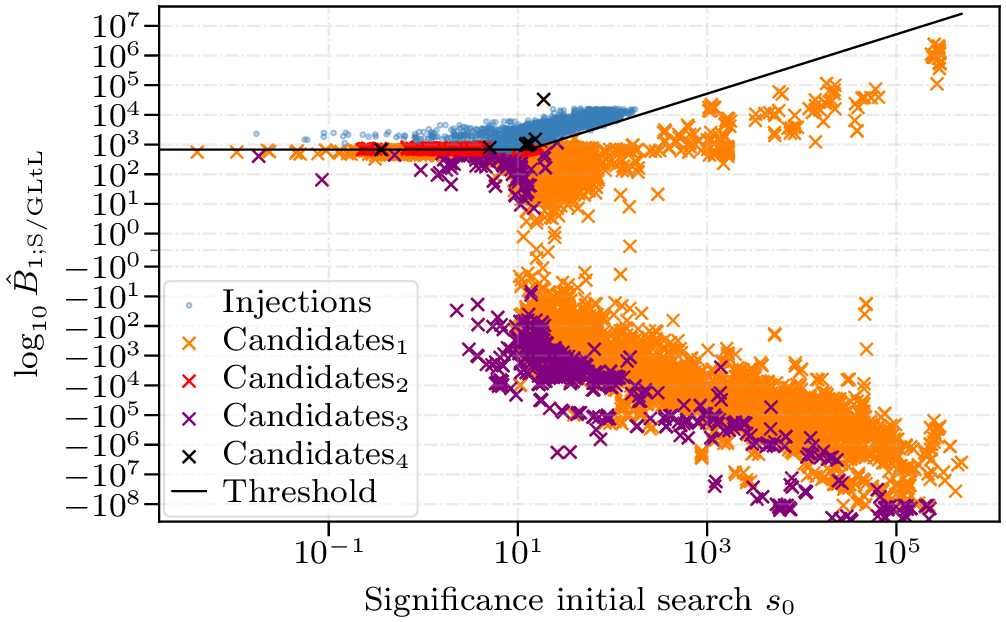}
    \caption{Distribution of significance values $s_1$ (left plot) and $\log_{10} \hat{B}_{1;\mathrm{S/GLtL}}$ (right plot) for the follow-up. In the right plot we use a ``symlog'' scale, i.e., linear between -1 and 1. 
    The blue circles show the $7\,200$ simulated test-signals, and the crosses indicate the $127\,500$ search candidates, with black crosses marking candidates that pass both vetoes, red crosses candidates that pass the $\log_{10} \hat{B}_{1;\mathrm{S/GLtL}}$ veto but not the significance veto, and purple crosses candidates that pass the significance veto but not the $\log_{10} \hat{B}_{1;\mathrm{S/GLtL}}$ veto}.
    The region below the black line is the candidate-rejection region based on the significance.
    \label{fig:FU}
  \end{center}
\end{figure*}

Figure~\ref{fig:FU} shows these quantities for both the simulated test-signals and the candidates. Based of the results from the simulated signals we define acceptance/rejection regions in the $s_0-s_1$ and $s_0-\log_{10} \hat{B}_\mathrm{1;S/GLtL}$ planes for the candidates. Our criteria
has a false dismissal probability of about $ 11 / 7\,200 \approx 0.15\%$ from 11 missed simulated signals and results in 15 candidates above both thresholds.

12 of these candidates can be associated with fake signals present in the data for validation purposes \citep[the so-called ``hardware injections", see][]{PhysRevD.95.062002}. The name ``hardware injection" stems from these signals being produced by moving the detectors' mirrors, rather than being added in software to the calibrated data. There are no hardware injection signals from the orbital parameter-space that we are investigating, but there are signals at the frequencies that we search from isolated neutron stars. Even though their waveforms do not match our binary ones, these signals are so loud, that our searches still identify them as unlikely to be due to noise. In particular 9 of our surviving candidates are related to the hardware injection at $\sim 52.8$\,Hz and 3 candidates to the one at $\sim 848.9$\,Hz.
The other 3 surviving candidates (at 507.9\,Hz, 517.2\,Hz, and 998.3\,Hz) can be associated with non-Gaussian disturbances, since they accumulate more than $80\%$ of their signal-to-noise ratio from very few segments. 
We conclude that none of these candidates can be associated with a defensible CW signal.

The main search has taken $\sim 10.15 \times 10^6$ CPU core-hours to complete, running on a combination of AMD EPYC 7402 CPUs and NVIDIA A100 GPUs. The computational cost of the follow-up is negligible.

\section{Estimated upper limits}
\label{sec:UL}

We estimate the 95\% confidence upper limits on the gravitational-wave amplitude $h_0^{95\%}$ in every 0.1\,Hz band. Our upper limit is the amplitude such that 95\% of a population of signals with frequency in that band and with the same distribution for the other parameters as the simulated test-signals (see appendix \ref{sec:robustness}) would have been detected.
The detection criterion for a simulated signal is that it generates a cluster with a higher significance than the 3rd cluster in its 0.1\,Hz band and its orbital parameter-space region.

Our estimate of the upper limit in a 0.1 Hz band centred at a frequency $f$ relies on the measurement of the noise amplitude spectral density $\sqrt{S_{\mathrm{n}}}$ in that band and on an estimate of the sensitivity depth ${\mathcal{D}}$ in that band.

The sensitivity depth connects the upper limit and the noise \citep{Behnke:2014tma,Dreissigacker:2018afk}:
\begin{equation}
\label{eq:sensDeph}
{\mathcal{D}}^{95\%}\equiv\frac{\sqrt{S_{\mathrm{n}}}}{h_0^{95\%}}.
\end{equation}
The expected value of the sensitivity depth, for results of the same search set-up, does not depend on frequency, and the noise spectral density from gravitational-wave detectors is also approximately constant over 0.1 Hz, unless the data is very disturbed.

We measure the upper limits in a few test-bands; using equation~\ref{eq:sensDeph}, we obtain different realisations of the sensitivity depth; we average them and obtain an estimate of the sensitivity depth; we use this to estimate the upper limits in all other bands, just based on the level of noise in each band.

Due to the different grid spacings of the initial stage below and above 500 Hz (see appendix~\ref{sec:details} for details), we determine two different sensitivity depth values, for the low and high frequency range of our search: $\left<{\mathcal{D}^{95\%}}_{\textrm{low}}\right>$ and $\left<{\mathcal{D}^{95\%}}_{\textrm{high}}\right>$. The average is taken over the $\mathcal{D}^{95\%}$ values obtained from the detection efficiency studies and upper limit measurements described above and in appendix \ref{sec:robustness}, carried out across the different orbital parameter-space regions and the relevant frequency test-bands.

This yields $\left<\mathcal{D}^{95\%}_{\textrm{low}}\right>= 18.1 \pm 1.1 ~\textrm{Hz}^{-1/2}$ and $\left<\mathcal{D}^{95\%}_{\textrm{high}}\right>=17.1 \pm 0.9~\textrm{Hz}^{-1/2}$, with the second number indicating the standard deviation.
We use these average sensitivity depths as a scale factor to determine the upper limits $h_0^{95\%}$ in all frequency bands using equation \ref{eq:sensDeph}. 

\begin{figure*}
  \begin{center}
    \includegraphics[width=1.0\columnwidth]{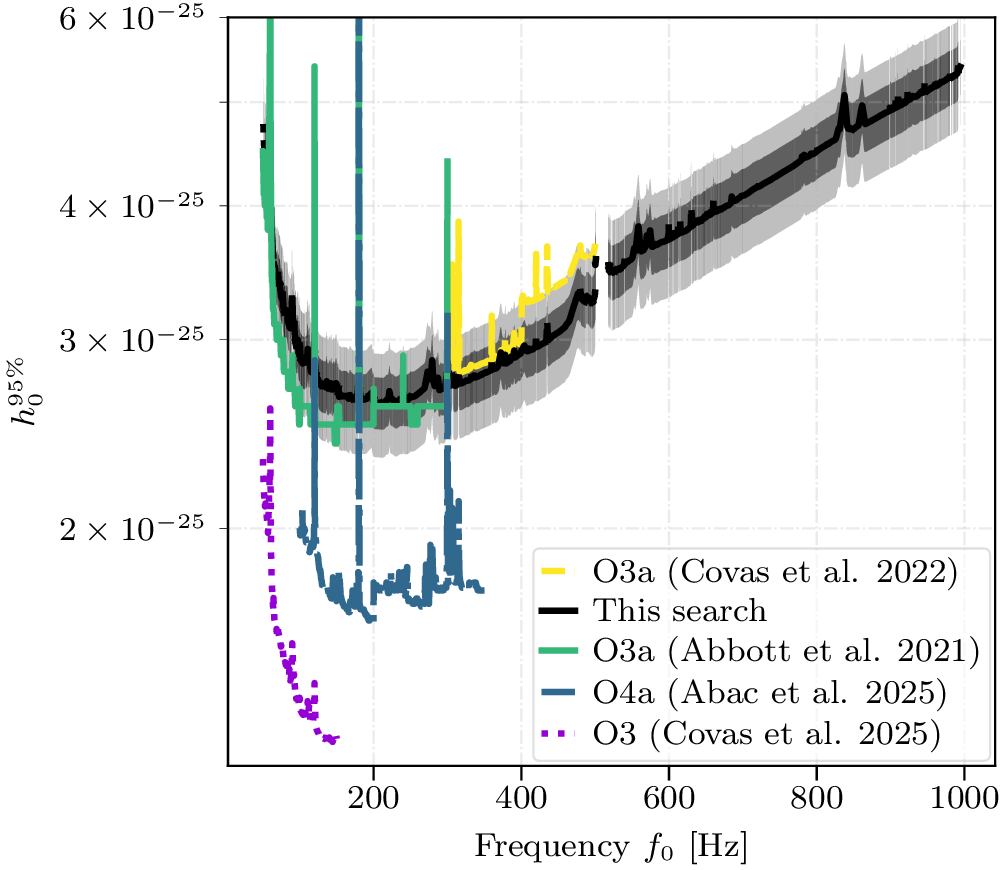}
    \includegraphics[width=1.0\columnwidth]{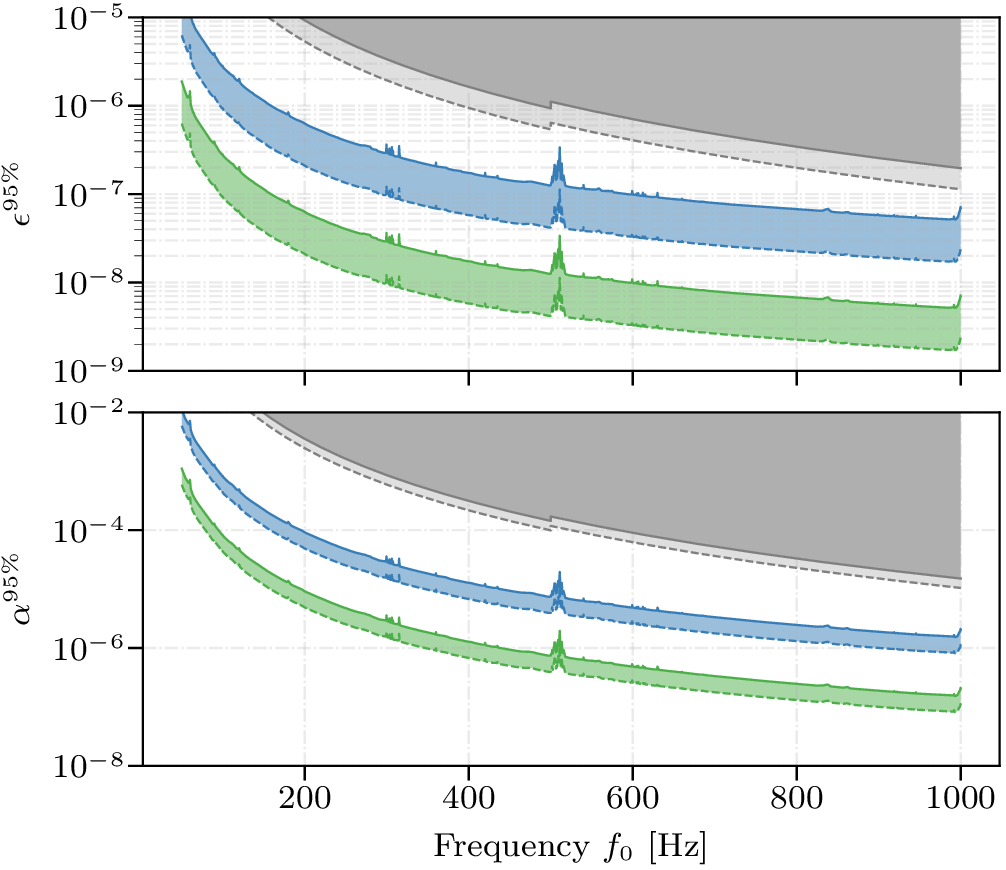}
    \caption{Estimated upper limits on the gravitational-wave amplitude $h_0^{95\%}$ (left plot), on the neutron star ellipticity $\epsilon$ (upper right plot), and on the r-mode amplitude $\alpha$ (lower right plot) at the $95\%$ confidence level as a function of the gravitational-wave frequency.
    Left plot: the black curve shows the results from this search, with the areas surrounding it indicating the $1\sigma$ and $2\sigma$ uncertainty regions. The other curves show the results from previous searches using O3 or O4 Advanced LIGO data.
    Right plots: the two different colors show results for distances of 100\,pc (upper blue curves) and 10\,pc (lower green curves).
    The upper shaded gray area shows the region where this search is not sensitive because the high ellipticities or r-mode amplitudes would generate a spin-down larger than the one probed by this search.
    In the upper right plot, the dashed lines assume $I_{zz} = 3 \times 10^{38}$\,kg\,m$^2$, while the solid lines assume $I_{zz} = 10^{38}$\,kg\,m$^2$.
    In the lower right plot, the dashed lines assume $M=1.9\,M_\odot$ and $R=13\,\mathrm{km}$, while the solid lines assume $M=1.4\,M_\odot$ and $R=11.7\,\mathrm{km}$.}
    \label{fig:UL}
  \end{center}
\end{figure*}

The estimated upper limits are shown in the left plot of figure~\ref{fig:UL}.
The lowest upper limit is $h_0^{95\%}=\hvalue$ at $f_0=\hfreq\,\mathrm{Hz}$.
These results are available in machine-readable format at \cite{Results1}. 

Because the sensitivity depth is estimated based on detection efficiency studies carried out in noise not obviously affected by disturbances, upper limits estimates in disturbed bands are not trustworthy. We identify disturbed bands based on the value of the maximum significance: when the measured maximum significance exceeds the expected maximum by more than 5 standard deviations, we assume that the band is disturbed and do not place an upper limit in it. The expected maximum significance is estimated by using a Gumbel distribution from extreme value theory. Out of 9\,500 0.1\,Hz bands we identify 1\,159 as disturbed.
These bands are given in machine-readable format at \cite{Results1}.

If the CWs are sourced by a mass asymmetry, the gravitational-wave amplitude is given by \citep{1996A&A...312..675B}
\begin{linenomath*}
\begin{equation}
\begin{aligned}
  h_0 \simeq 1.1 \times 10^{-26} & \left( \frac{f_0}{100\,\mathrm{Hz}} \right)^{2} \left( \frac{1\,\mathrm{kpc}}{d} \right) \\
                 & \times \left( \frac{\epsilon}{ 10^{-6} } \right) \left( \frac{ I_{zz} }{ 10^{38} \mathrm{~kg~m}^2} \right),
\end{aligned}
\label{eq:h0}
\end{equation}
\end{linenomath*}
where $\epsilon=|I_{xx} - I_{yy}|/I_{zz}$ is the equatorial ellipticity, $I_{zz}$ is the moment of inertia of the star with respect to the principal axis aligned with the rotational axis, and $d$ is the distance.
On the other hand, if the CWs are sourced by r-modes, the gravitational-wave amplitude is given by \citep{Owen_2010}
\begin{linenomath*}
\begin{equation}
\begin{aligned}
  h_0 & \simeq 1.0 \times 10^{-26} \left( \frac{f_0}{100\,\mathrm{Hz}} \right)^{3} \left( \frac{1\,\mathrm{kpc}}{d} \right) \\
                 & \times \left( \frac{\alpha}{2.8 \times 10^{-4}} \right) \left( \frac{M}{1.4 ~{\mathrm{M}}_\odot} \right) \left( \frac{R}{11.7 ~{\mathrm{km}}} \right)^3 \left( \frac{\tilde{J}}{0.0164} \right),
\end{aligned}
\label{eq:alpha}
\end{equation}
\end{linenomath*}
where $\alpha$ is the r-mode amplitude, $M$ is the mass of the neutron star, $R$ its radius, and $\tilde{J}$ a dimensionless parameter related to the equation of state.

The estimated upper limits on $h_0$ can hence be used to constrain the equatorial ellipticity and the r-mode amplitude of the targeted neutron star population by rearranging equations \ref{eq:h0} and \ref{eq:alpha}.
The ellipticity upper limits depend on the value for the moment of inertia of the neutron star, which is uncertain by around a factor of three \citetext{\citealp[see section 4B of][]{known_2007}\citealp[ and figure 4 of][]{refId0}}, although more exotic neutron star models such as quark stars, and lower-mass or larger-radius neutron stars could yield higher moments of inertia \citep{Horowitz_2010,Owen_2005}.
The r-mode amplitude upper limits depend on the neutron star mass and radius which are related by the neutron star equation of state.

The upper right plot of figure~\ref{fig:UL} shows the ellipticity upper limits for two distances and two values of the moment of inertia.
The tightest constraint for the ellipticity is achieved at $f_0 = 989.9$\,Hz, which for sources at 100\,pc with $I_{zz} = 10^{38}\,\mathrm{kg\,m^2}$ is $\epsilon < 5.2 \times 10^{-8}$.
If instead we assume $I_{zz} = 3 \times 10^{38}\,\mathrm{kg\,m^2}$, the upper limits are more stringent, as shown by the dashed traces.
For example, for sources within 100\,pc at 989.9\,Hz, we find $\epsilon < 1.7 \times 10^{-8}$, 
which is well within the predicted maximum ellipticities of neutron stars, that range between $10^{-8}$ and $10^{-5}$ \citep{mountainacc,maxquadrupolemass}.

The lower right plot of figure~\ref{fig:UL} shows the r-mode amplitude upper limits for two different distances and two mass-radius combinations.
The tightest constraint for the r-mode amplitude is achieved at $f_0 = 993.3$\,Hz, which for sources at 100\,pc with $M=1.4\,M_\odot$ and $R=11.7\,\mathrm{km}$ is $\alpha < 1.5 \times 10^{-6}$.
If instead we assume $M=1.9\,M_\odot$ and $R=13\,\mathrm{km}$, our constraint becomes more stringent: $\alpha < 8.1 \times 10^{-7}$.
Theoretical predictions on r-mode amplitudes are $\alpha\sim 8\times10^{-7}-10^{-4}$ \citep{Gusakov:2013jwa,gusakov_explaining_2014} or $\alpha\sim10^{-5}$ \citep{bondarescu_spin_2007}, which are by and large within the range that we have probed.

\section{Conclusions}
\label{sec:conclusions}

In this paper we present the most extensive search to date for CWs from unknown neutron stars in binary systems, with gravitational-wave frequencies between 50 and 1\,000\,Hz and orbital periods between 0.2 and 150\,days.
The main objective of this search is to cover the widest possible parameter-space region with a certain computational budget. 
For this reason we use coarser template grids and a shorter coherent time-baseline compared to our other recent search \citep{Covas_2025}, whose main objective was to achieve the best possible sensitivity.

We use the \textsc{BinarySkyHou$\mathcal{F}$} pipeline, which is the most efficient and sensitive pipeline to carry out this type of search. We do not detect any astrophysical signal. 

Previous searches such as \cite{O3aallskybinaryLIGO,theligoscientificcollaboration2025allskysearchcontinuousgravitationalwave} are $\approx$ 20\% more sensitive than this search but cover a much smaller parameter-space. This search has a similar sensitivity to our previous search \citep{Covas_2022}, but also covers a much larger parameter space region. Specifically this search reaches frequencies $\sim 3$ times larger, which is particularly difficult because the computational cost scales $\propto f_0^5$. Additionally this search covers large regions of the orbital parameter-space, some of which have not been searched before. Using the ``breadth" $\mathcal{B}$ to quantify the extent of the investigated parameter space, from equation 75 of \cite{WETTE2023102880}, we find that our parameter space is about 4 orders of magnitude larger than ever investigated with Advanced LIGO data. Our previous search \citep{Covas_2025} has around twice better sensitivity, but again covering a much smaller region of parameter space.

As the plots of figure~\ref{fig:ParameterSpace} show, vast regions of parameter space remain unexplored. This search demonstrates what state-of-the art search techniques can achieve in the challenging high-frequency range and may be used as blueprint for investigating new and more sensitive datasets.

\begin{acknowledgments}

We acknowledge the contribution of Benjamin Steltner who prepared the SFT data for this search.

This project has received funding from the European Union's Horizon 2020 research and innovation program under the Marie Sklodowska-Curie Grant Agreement No. 101029058.

This work was supported by the Universitat de les Illes Balears (UIB) with funds from the Programa de Foment de la Recerca i la Innovació de la UIB 2024-2026 (supported by the yearly plan of the Tourist Stay Tax ITS2023-086); the Spanish Agencia Estatal de Investigación (AEI) grants PID2024-157460NA-100, RED2024-153978-E, RED2024-153735-E, funded by MICIU/AEI/10.13039/501100011033 and the ERDF/EU; and the Comunitat Autònoma de les Illes Balears through the Conselleria d'Educació i Universitats with funds from the European Union - European Regional Development Fund (ERDF) (SINCO2022/18146 - Plataforma HiTech-IAC3-BIO).

We acknowledge PRACE for awarding us access to JUWELS Booster at GCS@JSC, Germany.

This work has utilized the ATLAS cluster computing at MPI for Gravitational Physics Hannover and the HPC system Raven at the Max Planck Computing and Data Facility.

This research has made use of data or software obtained from the Gravitational Wave Open Science Center (gwosc.org), a service of the LIGO Scientific Collaboration, the Virgo Collaboration, and KAGRA. This material is based upon work supported by NSF's LIGO Laboratory which is a major facility fully funded by the National Science Foundation, as well as the Science and Technology Facilities Council (STFC) of the United Kingdom, the Max-Planck-Society (MPS), and the State of Niedersachsen/Germany for support of the construction of Advanced LIGO and construction and operation of the GEO600 detector. Additional support for Advanced LIGO was provided by the Australian Research Council. Virgo is funded, through the European Gravitational Observatory (EGO), by the French Centre National de Recherche Scientifique (CNRS), the Italian Istituto Nazionale di Fisica Nucleare (INFN) and the Dutch Nikhef, with contributions by institutions from Belgium, Germany, Greece, Hungary, Ireland, Japan, Monaco, Poland, Portugal, Spain. KAGRA is supported by Ministry of Education, Culture, Sports, Science and Technology (MEXT), Japan Society for the Promotion of Science (JSPS) in Japan; National Research Foundation (NRF) and Ministry of Science and ICT (MSIT) in Korea; Academia Sinica (AS) and National Science and Technology Council (NSTC) in Taiwan \citep{GWOSC}.

\end{acknowledgments}

\appendix

\section{Initial stage set-up details}
\label{sec:details}

\begin{figure}[t]
  \begin{center}
    \includegraphics[width=1.0\columnwidth]{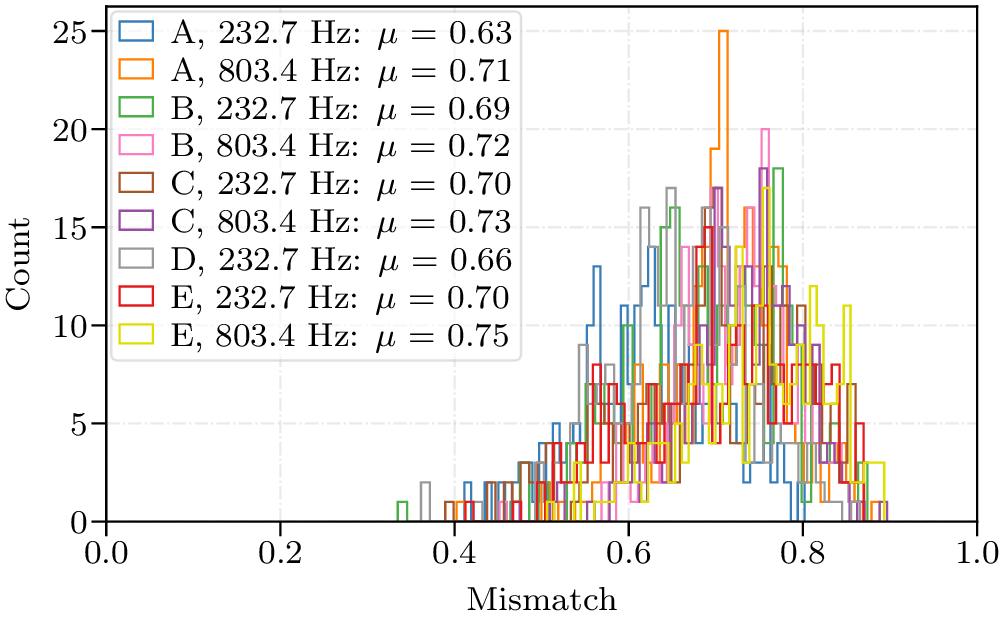}
    \caption{Mismatch distribution of this search at two different frequencies, in the 200 Hz and 800 Hz region, respectively. The other parameters that describe the CW signal span the ranges given in table~\ref{tab:region}.
    The legend shows the mean of each distribution for each different region of orbital parameter-space.}
    \label{fig:Mismatch}
  \end{center}
\end{figure}

\begin{figure}[t]
  \begin{center}
    \includegraphics[width=\columnwidth]{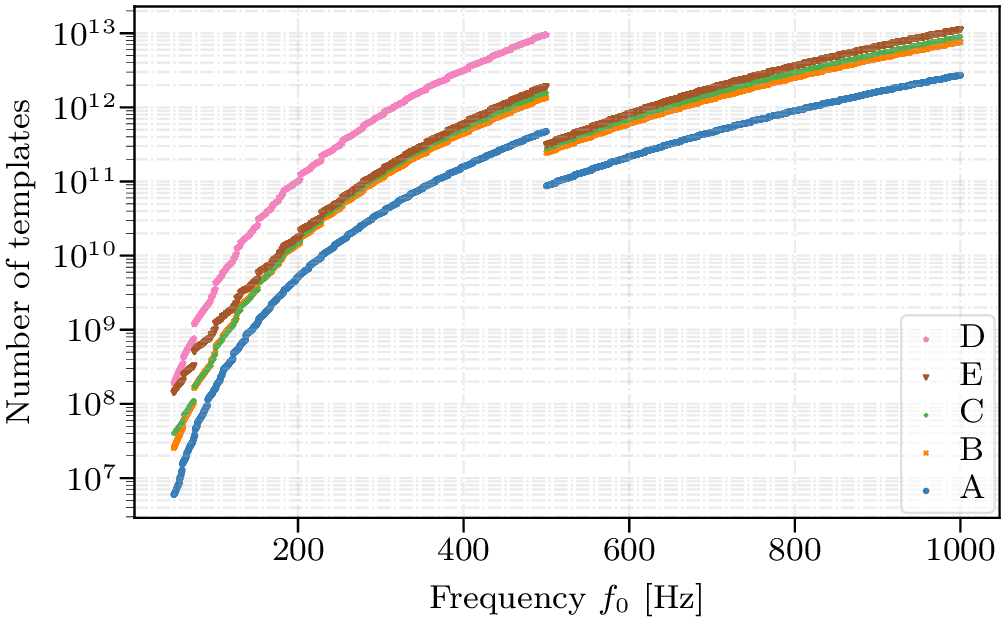}
    \caption{Number of templates searched in a 0.1\,Hz frequency band as a function of the gravitational-wave frequency.
    The different colors show the different regions of orbital parameter-space.
    The total number of templates is $\sim 6.3 \times 10^{16}$.}
    \label{fig:ntemplates}
  \end{center}
\end{figure}

The mismatch distributions for the various orbital parameter regions at two representative frequencies are shown in figure~\ref{fig:Mismatch}. The number of waveforms searched in each 0.1 Hz frequency band is shown in figure~\ref{fig:ntemplates}.

\section{Distribution of simulated test-signals and upper limits}
\label{sec:robustness}

The parameters of the simulated signals that we use to calibrate the follow-up procedure and to estimate the upper limits are drawn from the search ranges given in table~\ref{tab:region}. 
The sky position of the signals is drawn from a distribution isotropic on the celestial sphere.
The frequency $f_0$ is drawn uniformly from 0.1\,Hz test-bands. The noise in these test-bands is representative of the vast majority of the frequency ranges in our data, and is not affected by disturbances. We use 10 different bands below 500 Hz and 10 bands above 500 Hz.
Spin-down $f_1$ and eccentricity $e$ are log-uniformly distributed. The smallest spin-down and eccentricity values are taken two orders of magnitude below the maximum values.
The amplitudes $h_0$ are chosen by uniformly sampling the sensitivity depth given by equation~\ref{eq:sensDeph} from the range $[12, 49.5]$\,Hz$^{-1/2}$ (in discrete steps of $2.5$\,Hz$^{-1/2}$, resulting in 16 sensitivity depth values).
The other amplitude parameters $\cos\iota$, $\psi$, and $\phi_0$ are uniformly distributed.

To estimate the average sensitivity depth values that appear in equation \ref{eq:sensDeph} we use 32\,000 signals for each orbital region (16\,000 for region D due to the reduced frequency range), with 100 signals per frequency band and sensitivity depth value.

To determine the follow-up thresholds shown in figure~\ref{fig:FU}, we use a subset of test-signals detected by the initial stage, i.e., 1\,600 signals for each orbital region (800 for region D due to the reduced frequency range), corresponding to 10 signals for each frequency band and sensitivity depth value (using 5 of the 10 frequency bands below 500 Hz and 5 of the 10 bands above 500 Hz). We do this to limit the computational cost of these tuning studies.

\begin{figure}[b]
  \begin{center}
    \includegraphics[width=1.0\columnwidth]{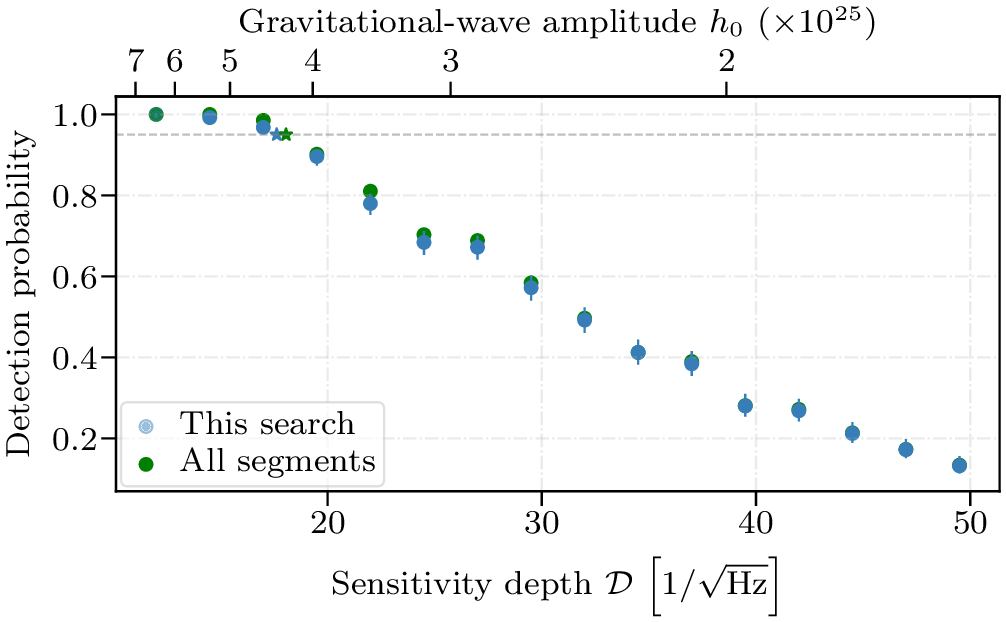}
    \caption{Illustrative example of the detection probability as a function of the normalized inverse signal-amplitude ($\mathcal{D}$) for signals at  $\approx$ 803.4\,Hz and with orbital parameters from region A. The top axis shows the corresponding values of the signal amplitude ($h_0$). The error bars show the $1\sigma$ uncertainty, and the star markers show the interpolated value of $\mathcal{D}^{95\%}$. Blue color shows results of this search, while green shows results when all segments are used in the first step of the initial stage.}
    \label{fig:sensdepth}
  \end{center}
\end{figure}

To estimate the upper limits we compute the fraction of detected signals as a function of the sensitivity depth $\mathcal{D}$ of the signals, a standard procedure also employed by several other searches \citep{O3aallskybinaryLIGO,O2aallskybinary,Steltner:2020hfd}.
We determine the sensitivity depth necessary to achieve 95\% detection efficiency for each frequency band and orbital region combination. We do this by using an interpolation, as exemplified by figure~\ref{fig:sensdepth}, which also shows the results obtained when no segments are discarded at the first step of the initial search, confirming that $\mathcal{D}^{95\%}$ is increased by less than $5\%$. We finally average the 95\% confidence sensitivity depth values from the low-frequency bands and from the high-frequency bands, yielding $\left<\mathcal{D}^{95\%}_{\textrm{low}}\right>=18.1$ $\textrm{Hz}^{-1/2}$ and $\left<\mathcal{D}^{95\%}_{\textrm{high}}\right>=17.1 $ $\textrm{Hz}^{-1/2}$, respectively.

We have only tested the sensitivity of our search with signals distributed in the ranges given in table~\ref{tab:region}.
If a putative signal lies outside these ranges, then the false-dismissal probability that we have estimated would be increased.
Quantifying this is outside of the scope of this paper.

The signal model described in section~\ref{sec:signal} and commonly used in CW searches assumes phase coherence for the duration of each coherent segment.
While this model is a very reasonable assumption for fast spinning neutron stars, it is known that some pulsars exhibit glitches \citep{PhysRevD.96.063004}, spin-wandering \citep{spinwandering}, and timing noise \citep{PhysRevD.91.062009}.
A loss of phase coherence will impact the sensitivity of a search if it is resolvable by it.
We refer the reader to the detailed discussion of section 4.2 in \citet{Covas_2025} regarding the robustness of our results to deviations from the assumed CW signal model.
We remark that the longest coherent time that we have used in this search is only 2\,700 s, shorter than the one used in that search, making these results more robust to deviations of the signal from the assumed model.

\bibliography{Paper}{}
\bibliographystyle{aasjournal}

\end{document}